\documentstyle[amssymb,preprint,aps]{revtex}

\begin{document}
\title{Multi-photon resonances in pure multiple-pulse NQR }
\author{G. B. Furman, G.E Kibrik$^{\ast }$, and A.Yu. Polyakov$^{\ast }$}
\address{Department of Physics, Ben-Gurion University,\\
Be'er-Sheva, Israel}
\author{G.E. Kibrik and A.Yu. Polyakov}
\address{Department of Physics, Perm State University,\\
Perm, Russia}
\date{\today }
\maketitle

\begin{abstract}
We have observed multi-photon resonances in a system with a spin 3/2
irradiated simultaneously by a multiple pulse radiofrequency sequence and a
low frequency field swept in the range 0-80 kHz. The used excitation scheme
allowed us to measure the effective field of the radiofrequency sequence. A
peculiarity of this scheme is that the intensity of the resonance lines
decreases slowly with the mode number. The theoretical description of the
effect is presented using both the rotating frame approximation and the
Floquet theory. Both approaches give indentical results at the calculation
of the resonance frequencies, transition probabilities and shifts of
resonance frequency. The calculated magnetization vs. the frequency of the
low-frequency field agrees well with the obtained experimental data. The
multi-frequency spectra give a way for studying slow atomic motion in solids.
\end{abstract}

{\bf Introduction}

One of the most effective and promising high-resolution nuclear magnetic
resonance (NMR) and nuclear quadrupole resonance (NQR) techniques for the
study of solids is a multiple-pulse radiofrequency (RF) action \cite%
{U.Haeberlen,M.Mehring}. The multiple-pulse methods allow one to remove
dipolar broadening from a resonance line in solids thus increasing by
several orders the sensitivity of the NMR and NQR spectroscopy in the study
of weak interaction.\ These methods are very effective in the study of the
spin lattice relaxation processes due to a slow atomic motion. Usually the
theoretical description of multiple-pulse experiments both in NMR \cite%
{Ivanov1978} and NQR \cite{Ainbinder} is based on the construction of the
effective time-independent Hamiltonian by using the conditions for
periodicity and cyclicity of the pulsed action . Then the dynamics of a spin
system subjected by pulsed RF fields is presented in an equivalent form as
the motion of nuclear spins in a constant effective field $H_{e}$ \cite%
{M.Maricq1986}. The magnitude and direction of this effective field are
determined by parameters of the multiple-pulse sequence. An experimental
measurement of the value of the effective field is important for the
confirmation of this theoretical model.

It is reasonable to suggest that an additional field with an angular
frequency $\Omega $ close to $\omega _{e}=\gamma H_{e}$ should cause
resonance absorption of energy ($\gamma $ is the gyromagnetic ratio of
nuclei).\ Spin-echo signals observed between RF pulse sequence would allow
us to determine $H_{e}$ as well as\ to obtain the information on slow atomic
motion that is not available from the traditional metrods

With this in mind,\ we have studied experimentally resonance transitions in
the nuclear spin system subjected by a simultaneous action of a
multiple--pulse RF sequence and an additional low frequency (LF) field with
an angular frequency $\Omega $. The results of our experiments described in
the next section have shown that resonance transitions were observed not
only at the frequency close to $\Omega _{0}=\omega _{e}$, but also at
frequencies close to $\Omega _{n}$ given by the expression: 
\begin{equation}
\Omega _{n}=\left| \omega _{e}\pm 2\pi n/t_{c}\right| ,\text{ \ \ \ \ \ \ \
\ }n=\pm 1,\pm 2...  \eqnum{1}
\end{equation}
where $t_{c}$ is the period of the multiple pulse RF sequence.

Multiple resonance modes of higher orders have been detected by microwave
spectroscopy\cite{A.Autler1955}, molecular beam technique \cite{H.Salwen},
optical pumping\cite{Margerie1955} and observed previously in NMR\
experiments\cite{P.Bucci,Y.Zur,Vega1992,Demco1999,Demco}. However the
amplitude of these resonances decreased abruptly with the mode order of the
resonance.\ As distinct from this, the amplitude of the resonances observed
in our experiments decreased slowly and the resonances of higher orders were
well observable.\ 

Because the nuclear spin system possesses a set of the resonance
frequencies, the relaxation measurements performed on one resonance
frequency $\Omega _{n}$\ can give the information on oscillations of atoms
on all the frequencies from the spectrum determined by (1). It moves us to
comprehensive experimental and theoretical study of this system.

The theoretical treatment of NMR phenomena is usually based on three
approaches: i) a semi - classical mathematical approach \cite{A.Abragam};
ii) a second quantization method \cite{CohenTann}; and iii) the Floquet
theory \cite{J.H.Shirley}.

The semi - classical mathematical approach \cite{A.Abragam}, where the field
is considered as a classical system and the atomic system as a quantum one,
has allowed one to explain of series of experimentally observed phenomena.
This approach is quite natural if to take into account, that the average
number of photons in a mode of the periodic field is extremely great. The
main method used in the framework of the semi - classical approach is the
so-called ''rotating frame approximation'', keeping exactly just the terms
that are resonant. The remaining non-resonant terms are considered as a
perturbation.

Intrinsic inconsistency of the semi-classical approach is obviated in the
framework of the secondary quantization method \cite{CohenTann,C.Cohen1994}.
Treating the RF field as photons, the evolution of the united system
\textquotedblright atom +field\textquotedblright\ (so called
\textquotedblright dressed\textquotedblright\ atom) is described by the
Hamiltonian which is independent of time, and its investigation turns out
simpler than solving the Schr\"{o}dinger equation with the time-dependent
Hamiltonian. With the time-independent Hamiltonian, one can define energy
levels of the physical system. Each of these levels corresponds to a
stationary state of the system:\ atom dressed by photons. The dependence of
the energy of the united system on the field frequency allows one to
construct the diagram of levels. The coupling between RF photons and an atom
perturbs the energy levels and allows one to interpret all resonances as
\textquotedblright crossings\textquotedblright\ and \textquotedblright
anticrossings\textquotedblright\ of the conforming levels \cite{Cohen1973}.

The Floquet theory \cite{J.H.Shirley} is a power method widely used in NMR
spectroscopy for solving time-dependent problems \cite{Y.Zur,Vega1986}. On
the one hand, this approach uses the advantages of the secondary
quantization method resulting in the time-independent Hamiltonian. On the
other hand, it simplifies calculations as the terms related to the system of
free RF photons are not included in the Hamiltonian . The last is justified
because an RF field applied in the NMR method contains abundant photons and
changes in the state of the photon system may be neglected. In the framework
of the Floquet theory, the interaction of the spin system with RF field is
considered as completely quantum-mechanical. Therefore, the Floquet theory
can be considered to be a bridge between the semi-classical and the quantum
methods.

To calculate the transition probabilities and the shift of the center of the
resonance line, two theoretical methods are used: the semi-classical
approach and the Floquet theory. We extend the theory of the multiphoton
resonances \cite{A.Abragam,CohenTann,J.H.Shirley} to the case of the pure
multiple-pulse NQR. The results given by both theoretical approaches are
compared with each other and with the experiment.

{\bf Experiments}

The experiments were performed using an automated multiple-pulse NQR
spectrometer. Mutually perpendicular continuous LF and pulse RF magnetic
fields were generated by crossed coils. Resonances were observed in the
effective field of a multiple-pulse RF sequence $\left( \pi /2\right)
_{y}-\left( t_{c}/2-\varphi -t_{c}/2\right) ^{N}$ , where $\varphi $ denotes
the pulse that rotates the nuclear magnetization about direction of RF field
in the rotating frame by an angle $\varphi $, and $N$ is the number of
pulses in the sequence. This sequence consisted of \ $N=256$ pulses and spin
locking signal was sampled in the interval between them. The period of the
pulse sequence was $t_{c}=100ms$.

The NQR of $^{35}Cl$ nuclei was observed in polycrystalline $KClO_{3}$ at $%
77 $ $K$. Without action of the LF field, we recorded the sequence of echo
signals in the interval between pulses at $28.9539$ $MHz$ with practically
constant amplitude (height of the spin echo signals) during the observation
time (Fig. 1). This amplitude corresponds to a quasi-equilibrium state of
the spin system with magnetization $M_{0}$.

LF field with the amplitude $2.5$ $G$, nonsynchronized with RF pulses, was
swept in the range $1\div 80$ $KHz$. The resonance reduction of the
magnetization amplitude was observed (Fig.1) at several different
frequencies $\Omega $ .\ \ The transient process with the spin-spin
relaxation time $T_{2}$ was followed by establishment of a new
quasi-equilibrium state of the spin system with the reduced amplitude of the
magnetization. {\Large \ }Our measurements give the value $T_{2}$=455 $ms$%
{\Large .}

Fig. 2 shows the dependence of the relative magnetization $M/M_{0}$ on the
frequency $\Omega $ of the LF field for the $\varphi =\frac{\pi }{2}$ RF
pulse (corresponding to the pulse duration{\large \ }$t_{w}=15${\large \ }$%
\mu s${\large \ }). The effective frequency of the RF field is determined by 
$\omega _{e}=$ $\frac{\varphi }{t_{c}}$ . For example, $\omega _{e}$\ $=2.5$ 
$kHz$\ for $\varphi =\frac{\pi }{2}$.

As follows from Fig. 2, the amplitude of the resonances decreases slowly
with increasing the mode number $n$.

{\bf Theory}

{\bf 1. Semi-classical approximation: rotating frame approximation}

Let us consider a system of nuclear spins $I=3/2$ placed in an inhomogenous
electric field gradiet (EFG) and subjected to a joint action of two
time-dependent magnetic fields: a multiple--pulse RF sequence with the
angular frequency $\omega $ equaled to the resonance frequency $\omega _{0}$
and a continuous low frequency (LF) field with the angular frequency $\Omega 
$.

In the rotating frame the equation for the density matrix of the system
takes the form: 
\begin{equation}
i\frac{d\rho \left( t\right) }{dt}=\left[ {\cal H}\left( t\right) ,\rho
\left( t\right) \right] ,  \eqnum{2}
\end{equation}%
where the Hamiltonian of the system is 
\begin{equation}
{\cal H}\left( t\right) =2\omega _{2}S_{z}\cos \Omega t+S_{x}\varphi
\sum_{k=1}^{\infty }\delta \left( t-kt_{c}-t_{c}/2\right) ,  \eqnum{3}
\end{equation}%
$S_{x}$ and $S_{z}$ are $X$- and $Z$-components of the effective spin
operator \cite{Ainbinder}; the pulse angle $\varphi =\ \gamma H_{1}t_{w}$
and $0<\varphi <2\pi $; $\omega _{2}=\gamma H_{2}$; $H_{1}$ and $H_{2}$ are
the amplitudes of the RF pulse and LF fields, respectively.\ The initial
phase of LF field is chosen equal zero.

To solve Eq.(2), we apply the unitary transformation 
\begin{equation}
\tilde{\rho}\left( t\right) =e^{-i\Omega S_{x}t}P^{+}\left( t\right) \rho
\left( t\right) P\left( t\right) e^{i\Omega S_{x}t},  \eqnum{4}
\end{equation}
where the unitary operator $P\left( t\right) $ is given by the solution of
the equation 
\begin{equation}
i\frac{dP\left( t\right) }{dt}=S_{x}\varphi \sum_{k=1}^{\infty }\delta
\left( t-kt_{c}-t_{c}/2\right) P\left( t\right) -P\left( t\right) {\cal H}%
_{e}  \eqnum{5}
\end{equation}
with the initial condition 
\begin{equation}
P\left( 0\right) =1  \eqnum{6}
\end{equation}
and the effective time-independent Hamiltonian 
\begin{equation}
{\cal H}_{e}=\omega _{e}S_{x}.  \eqnum{7}
\end{equation}

After performing the unitary transformation (4) we obtain the following
equation for the transformed density matrix $\tilde{\rho}\left( t\right) $ 
\begin{equation}
\frac{d\tilde{\rho}\left( t\right) }{dt}=\left[ {\cal H}^{tr}\left( t\right)
,\tilde{\rho}\left( t\right) \right]  \eqnum{8}
\end{equation}
with 
\begin{equation}
{\cal H}^{tr}\left( t\right) =\left( \omega _{e}-\Omega \right) S_{x}+\tilde{%
S}_{z}\left( t\right) ,  \eqnum{9}
\end{equation}
where $\tilde{S}_{z}\left( t\right) =P^{+}\left( t\right) S_{z}P\left(
t\right) $ is the periodic function of time and can be expanded into the
Fourier series 
\begin{equation}
\tilde{S}_{z}\left( t\right) =S_{+}\sum_{n=-\infty }^{\infty
}b_{n}e^{-in\omega _{c}t}+S_{-}\sum_{n=-\infty }^{\infty }b_{n}e^{in\omega
_{c}t}  \eqnum{10}
\end{equation}
where $S_{\pm }=S_{z}\pm iS_{y}$, $b_{n}=\omega _{2}\frac{\left( -1\right)
^{n}\sin \frac{\varphi }{2}}{\varphi +2\pi n}$ are the Fourier coefficients,
and $\omega _{c}=\frac{2\pi }{t_{c}}$.

The resonance terms in the solution of Eq. (8) are determined by the lowest
values of the differences $\xi =\left\vert \omega _{e}-\Omega \pm n\omega
_{c}\right\vert $. At $\xi =0$ we obtain the expression (1) for resonance
conditions. The rest non-resonant terms determine a\ frequency shift $\Delta 
$ for the position of the resonance dips shown in Figs 1 and 2. Let us
estimate this shift as a function of the frequency $\Omega $ using the
average Hamiltonian theory \cite{N.E.Ainbinder,M.Menabde1979}. We will
consider three various cases depending on the value of the frequency $\Omega 
$.

The first case, $\Omega =0$, corresponds to the appearance of an additional
constant magnetic field along $Z$-axes resulting in the difference between $%
\omega $ and $\omega _{0}$.

To determine the influence of all the non-resonant terms, we will perform
the unitary transformation under Eq.(8), 
\begin{equation}
\rho ^{\prime }\left( t\right) =e^{i\varphi S_{x}\frac{t}{t_{c}}}\tilde{\rho}%
\left( t\right) e^{-i\varphi S_{x}\frac{t}{t_{c}}},  \eqnum{11}
\end{equation}
which leads to 
\begin{equation}
\frac{d\rho ^{\prime }\left( t\right) }{dt}=\left[ {\cal H}^{\prime }\left(
t\right) ,\rho ^{\prime }\left( t\right) \right] =\left[ S_{+}\sum_{n=-%
\infty }^{\infty }b_{n}e^{-i\left( 2\pi n+\varphi \right) \frac{t}{t_{c}}%
}+S_{-}\sum_{n=-\infty }^{\infty }b_{n}e^{i\left( 2\pi n+\varphi \right) 
\frac{t}{t_{c}}},\rho ^{\prime }\left( t\right) \right] .  \eqnum{12}
\end{equation}
Let us replace $H^{\prime }\left( t\right) $ by the average Hamiltonian\ up
to second order\ of the expansion in $\omega _{2}/\omega _{e}$ : 
\begin{equation}
{\cal H}^{av}=t_{c}\sum_{n=-\infty }^{\infty }\frac{b_{n}^{2}}{2\left( 2\pi
n+\varphi \right) }\left[ S_{-},S_{+}\right] ,  \eqnum{13}
\end{equation}
Note that in the case of non-zero initial phase of the LF field $\bar{H}$\
includes $\left| b_{n}\right| ^{2}$. Therefore the initial phase of the LF
field does not influence the result. Performing\ the summation in Eq.(13) we
obtain

\begin{mathletters}
\begin{equation}
\Delta _{1}=\frac{\omega _{2}^{2}t_{c}}{8}\cot \frac{1}{2}\varphi \text{
with }\varphi \neq -2\pi n.  \eqnum{14}
\end{equation}

In the limit that $t_{c}\rightarrow 0$, a one long pulse, we obtain from Eq.
(14) a simple expression for the shift $\Delta _{1}=\frac{\omega _{2}^{2}}{%
4\omega _{e}}$, $\allowbreak $which is similar to the Bloch-Siegert results 
\cite{Bloch1940}. However, in contrast to the Bloch-Siegert shift which is
produced by the RF field and has the value of the order of\ $\left( \frac{%
\omega _{1}}{\omega _{0}}\right) ^{2}\sim 10^{-6}$, the obtained shift is
sufficiently higher because it is determined by the ratio of two weak fields
with the amplitudes $H_{2}$ and $H_{e}$ . For example, for the angle $%
\varphi =\frac{\pi }{2}$, the shift $\Delta _{1}$ is of the order of $%
10^{-2} $\ . The shift $\Delta _{1}$ is caused by a deviation of the
frequency of the applied RF field from the Larmor frequency and should be
taken into account when the results of multiple-pulse experiments are
analyzed.

The second case: $\Omega =\omega _{e}$ . The average Hamiltonian $\bar{H}$
is 
\end{mathletters}
\begin{equation}
{\cal H}^{av}{\cal =}2b_{0}S_{x}+\sum_{n=-\infty ,n\neq 0}^{\infty }\frac{%
b_{n}^{2}}{2n\omega _{c}}\left[ S_{-},S_{+}\right]  \eqnum{15}
\end{equation}
\qquad After the summation in Eq.(15) we obtain the shift due to common
action of the non-resonant terms 
\begin{equation}
\Delta _{2}=\frac{\omega _{2}^{2}t_{c}}{\varphi ^{3}}\left[ \sin ^{2}\frac{1%
}{2}\varphi -\frac{\varphi }{8}\left( \varphi +\sin \varphi \right) \right] .
\eqnum{16}
\end{equation}

Thus, the resonance frequency differs from$\ \omega _{e}$. A shift appears
also when we consider $\Omega =\omega _{e}+n\omega _{c}$ .

In order to determine correctly the resonance conditions, let us consider
the third case: $\Omega \neq \omega _{e}+n\omega _{c}$. Using the
transformation 
\begin{equation}
\rho ^{\prime \prime }\left( t\right) =e^{i\left( \varphi -\Phi \right) S_{x}%
\frac{t}{t_{c}}}\tilde{\rho}\left( t\right) e^{-i\left( \varphi -\Phi
\right) S_{x}\frac{t}{t_{c}}},  \eqnum{17}
\end{equation}
where $\Phi =\Omega t_{c}$, we obtain 
\begin{equation}
\frac{d\rho ^{\prime \prime }\left( t\right) }{dt}=\left[ S_{+}\sum_{n=-%
\infty }^{\infty }b_{n}e^{-i\left( \varphi -\Phi +2\pi n\right) \frac{t}{%
t_{c}}}+S_{-}\sum_{n=-\infty }^{\infty }b_{n}e^{i\left( \varphi -\Phi +2\pi
n\right) \frac{t}{t_{c}}},\rho ^{\prime \prime }\left( t\right) \right] . 
\eqnum{18}
\end{equation}
Let us exclude from the sums in (18) the terms with $n=-\frac{\varphi -\Phi 
}{2\pi }$ : 
\begin{equation}
\frac{d\rho ^{\prime \prime }\left( t\right) }{dt}=\left[ \left( -1\right) ^{%
\frac{\varphi -\Phi }{2\pi }}\frac{\omega _{2}}{\Phi }\sin \frac{\varphi }{2}%
S_{x}+S_{+}\sum_{n=-\infty ,n\neq -\frac{\varphi -\Phi }{2\pi }}^{\infty
}b_{n}e^{-i\left( \varphi -\Phi +2\pi n\right) \frac{t}{t_{c}}%
}+S_{-}\sum_{n=-\infty ,n\neq -\frac{\varphi -\Phi }{2\pi }}^{\infty
}b_{n}e^{i\left( \varphi -\Phi +2\pi n\right) \frac{t}{t_{c}}},\rho ^{\prime
\prime }\left( t\right) \right]  \eqnum{19}
\end{equation}
The average Hamiltonian\ up to second order\ of the expansion in $\omega
_{2}/\omega _{e}$ 
\begin{equation}
{\cal H}^{av}=\left( -1\right) ^{\frac{\varphi -\Phi }{2\pi }}\frac{\omega
_{2}}{\Phi }\sin \frac{\varphi }{2}S_{x}+\omega _{2}^{2}t_{c}\sin ^{2}\frac{%
\varphi }{2}\sum_{n=-\infty ,n\neq -\frac{\varphi -\Phi }{2\pi }}^{\infty }%
\frac{1}{\left( \varphi +2\pi n\right) ^{2}\left( \varphi -\Phi +2\pi
n\right) }\left[ S_{-},S_{+}\right] .  \eqnum{20}
\end{equation}
Because in the considered case $\varphi -\Phi \neq -2\pi n$, the first term
in (20) disappears and the summation over\ all $n$ gives the following
equation for the frequency shift: 
\begin{equation}
\Delta _{3}=\omega _{2}^{2}t_{c}\left( 1-\frac{2\sin \frac{\varphi }{2}\sin 
\frac{\Phi }{2}}{\Phi \sin \frac{\varphi -\Phi }{2}}\right) .  \eqnum{21}
\end{equation}

Emphasize that $\Phi $ is a function of $\Delta _{3}$ and Eq. (21) has
infinite number of roots. Resonance transitions are realized at $\Omega =%
\tilde{\Omega}_{n}$, where \ 
\begin{equation}
\tilde{\Omega}_{n}=\omega _{e}+\Delta _{3n}+n\omega _{c},\text{ }n=0,\pm
1,\pm 2,...  \eqnum{22}
\end{equation}
Resonance frequencies calculated according (22) using numerical solutions of
(21) for the shift are presented in Table 1 and 2 along with the
experimental data. The shift is caused by off-resonant component of the LF
field.

The time-average transition probability, $\bar{P}$ can be determined using
the method developed in \cite{L.Landau}\ : 
\begin{equation}
\bar{P}=\frac{1}{2}\sum_{n}\frac{b_{n}^{2}}{\left( \omega _{e}-\Omega
-n\omega _{c}+\Delta _{3}\right) ^{2}+b_{n}^{2}},  \eqnum{23}
\end{equation}
Since $\left( \frac{\omega _{2}}{\omega _{e}}\right) ^{2}<<1$\ (for example, 
$\left( \frac{\omega _{2}}{\omega _{e}}\right) ^{2}=6.\,\allowbreak 5\times
10^{-2}$\ for $\varphi =\frac{\pi }{2}$), the transition probabilities
become significant only in the vinicity of the resonance frequency
determined by the condition (22). The expression for the nuclear
magnetization $M$ , the quantity needed to compare with experiment, can be
found using Eq. (23): 
\begin{equation}
\frac{M}{M_{0}}=1-\sum_{n}\frac{b_{n}^{2}}{\left( \omega _{e}-\Omega
-n\omega _{c}+\Delta _{3}\right) ^{2}+b_{n}^{2}}.  \eqnum{24}
\end{equation}
Expressions (23) and (24), obtained in the framework of the semi-classical
approach, determine a multi-line resonance spectrum of the two-energy level
system with spin 1/2 irradiated by multiple pulse RF and LF fields. As
follows from (24) the amplitudes of the resonances decreases slightly with
the number $n$. There is disparity between a two-level energy model of the
spin system and the multi-line resonance spectrum. In order to make a
physical interpretation clear, the energy spectrum and transition
probabilities will be calculated in the next section by using the Floquet
theory\cite{J.H.Shirley}.

\bigskip

{\bf 2. Floquet theory}

The Floquet theory provides a quite natural description of the multi-line
resonance spectrum of the two-energy level spin system. Let us introduce the
time-dependent operator 
\begin{equation}
U\left( t,t_{0}\right) =F\left( t\right) F^{+}\left( t_{0}\right) , 
\eqnum{25}
\end{equation}
where the unitary operator $F\left( t\right) $ obeys to the evolution
equation 
\begin{equation}
\frac{dF\left( t\right) }{dt}={\cal H}^{tr}\left( t\right) F\left( t\right) 
\eqnum{26}
\end{equation}
with the periodic time-dependent Hamiltonian (9). The general form of the
solution of Eq.(26) is given by the Floquet theorem \cite{J.H.Shirley} is 
\begin{equation}
F\left( t\right) =\Psi \left( t\right) e^{-i{\cal H}^{F}t},  \eqnum{27}
\end{equation}
where $\Psi \left( t\right) $ is a periodic function of time with the
frequency $\omega _{c}$, and ${\cal H}^{F}$ is the time-independent
effective Hamiltonian Using the periodic character of the Hamiltonian $%
\tilde{H}\left( t\right) $ and function $\Psi \left( t\right) $, the
time-dependent Eq. (26) can be rewritten as an infinite set of the coupled
equations \cite{J.H.Shirley} 
\begin{equation}
\sum_{\mu k}\left( {\cal H}_{\alpha \mu }^{n-k}+n\omega _{c}\delta
_{nk}\delta _{\alpha \mu }\right) F_{\mu \beta }^{k}=h_{\beta }F_{\alpha
\beta }^{n},  \eqnum{28}
\end{equation}
where $h_{\beta }$ are diagonal elements of the effective Hamiltonian ${\cal %
H}^{F}$ , ${\cal H}_{\alpha \mu }^{n-k}$ and $F_{\alpha \beta }^{n}$ are the
Fourier components of the matrix elements of ${\cal H}^{t}\left( t\right) $
and $F\left( t\right) $: 
\begin{equation}
F_{\alpha \beta }\left( t\right) =\sum_{n}F_{\alpha \beta }^{n}e^{in\omega
_{c}t}e^{-ih_{\beta }},  \eqnum{29}
\end{equation}
\begin{equation}
{\cal H}_{\alpha \beta }^{t}\left( t\right) =\sum_{n}{\cal H}_{\alpha \beta
}^{n}e^{in\omega _{c}t}.  \eqnum{30}
\end{equation}
Here $\alpha =\pm 1/2$, $\beta =\mp 1/2$ characterize various spin states.
Using the orthogonal basis $\left| \alpha n\right\rangle $ , where $n$
denotes the Fourier component, the term in the brackets of the Eq. (28) can
be presented as the Floquet Hamiltonian \cite{J.H.Shirley} 
\begin{equation}
\left\langle \alpha n\right| {\cal H}_{F}\left| \beta k\right\rangle ={\cal H%
}_{\alpha \beta }^{n-k}+n\omega _{c}\delta _{nk}\delta _{\alpha \beta }, 
\eqnum{31}
\end{equation}

Using the operator basis $e_{an,\beta m}$ \cite{N.E.Ainbinder} with matrix
elements $\left\langle \mu p\right| e_{an,\beta m}\left| \eta q\right\rangle
=\delta _{\alpha \mu }\delta _{\beta \eta }\delta _{pn}\delta _{mq}$ the
Floquet Hamiltonian can be expressed as 
\begin{equation}
{\cal H}_{F}=\sum_{\alpha ,n}\left( \varepsilon _{\alpha }+n\omega
_{c}\right) e_{\alpha n,\alpha n}+\sum_{\alpha \neq \beta ,n,m}\left(
b_{m-n}^{\alpha \beta }e_{\alpha n,\beta m}+b_{n-m}^{\beta \alpha }e_{\beta
n,\alpha m}\right)  \eqnum{32}
\end{equation}
where $\varepsilon _{\alpha }=\alpha \left( \omega _{e}-\Omega \right) ,\
b_{m-n}^{\alpha \beta }=b_{m-n}$ are the Fourier coefficients determined in
(10). The first term in (32) is diagonal, while the second one is
off-diagonal.

As follows from (32), nonzero off-diagonal matrix elements between $%
\left\vert \alpha n\right\rangle $ and $\left\vert \beta m\right\rangle $
for various $n$ and $m$ appear in first order of $\frac{\omega _{2}}{\omega
_{e}}$. Therefore all resonance transitions between different states of the
Hamiltonian (32) can be excited in the straight way without involving
intermediate states as it is used in the scheme of multiphoton excitation
proposed in \cite{Y.Zur,Vega1986,Gromov} .

The intensity of a resonance transition can be estimated using the
perturbation method \cite{H.Salwen} based on the approximation of the
Floquet Hamiltonian $H_{F}$ (32) by the $2\times 2$ matrix $\left( \alpha
\neq \beta \right) $ 
\begin{equation}
{\cal H}_{F}^{\prime }= 
\begin{tabular}{ll}
$\varepsilon _{\alpha }+m\omega _{c}+c_{\alpha m}$ & $b_{m-n}+d_{n-m}^{%
\alpha \beta }$ \\ 
$b_{m-n}+d_{n-m}^{\alpha \beta }$ & $\varepsilon _{\beta }+n\omega
_{c}+c_{\beta n}$%
\end{tabular}
,  \eqnum{33}
\end{equation}
with matrix elements corrected by taking into account the rest part of the
Hamiltonian: the diagonal elements - by introducing the additional terms $%
c_{\alpha n}$\ which are up to second order to within $\left( \frac{\omega
_{2}}{\omega _{e}}\right) ^{2}\ll 1$\ is 
\begin{equation}
c_{\alpha n}=\sum_{\mu ,k}\frac{\left\langle \alpha n\right| {\cal H}%
_{F}\left| \mu k\right\rangle \left\langle \mu k\right| {\cal H}_{F}\left|
\alpha n\right\rangle }{E_{\alpha n}-E_{\mu k}}.  \eqnum{34}
\end{equation}
and \ the off-diagonal ones - by adding the terms\ $d_{n-m}^{\alpha \beta }$
of the same order

\begin{equation}
d_{n-m}^{\alpha \beta }=2\sum_{\mu k}\frac{\left\langle \beta n\right\vert 
{\cal H}_{F}\left\vert \mu k\right\rangle \left\langle \mu k\right\vert 
{\cal H}_{F}\left\vert \alpha m\right\rangle .}{E_{\beta n}-E_{\mu k}}, 
\eqnum{35}
\end{equation}%
In (34) and (35) $E_{\beta n}$ are eigenvalues of the Floquet Hamiltonian $%
H_{F}$. Diagonalizing the matrix (33) gives the eigenvalues of the
Hamiltonian $H_{F}^{\prime }$: 
\begin{eqnarray}
2E_{m,n}^{\pm \left( \alpha \beta \right) } &=&\left[ \varepsilon _{\alpha
}+\varepsilon _{\beta }+\left( m+n\right) \omega _{c}\right] \pm  \nonumber
\\
&&\pm \left\{ \left[ \left( \varepsilon _{\alpha }+m\omega _{c}+c_{\alpha
m}\right) -\left( \varepsilon _{\beta }+n\omega _{c}+c_{\beta n}\right) %
\right] ^{2}+\left( b_{m-n}+d_{n-m}^{\alpha \beta }\right) ^{2}\right\}
^{1/2}.  \eqnum{36}
\end{eqnarray}%
Substituting the matrix elements (31) into (34) we obtain the expressions
for the energy level shifts 
\begin{equation}
c_{\alpha n}=-c_{\beta n}=\sum_{k=-\infty }^{\infty }\frac{b_{k-n}^{2}}{%
\left( \varphi -\Phi +2\pi \left( k-n\right) \right) }.  \eqnum{37}
\end{equation}%
The summation in Eq.(37) gives the same equation for the frequency shift as
above obtained using the rotating frame approximation (see Eq.(20)): 
\begin{equation}
\Delta _{3}=c_{\alpha n}-c_{\beta n}=\frac{\omega _{2}^{2}t_{c}}{2}\left( 1-%
\frac{2\sin \frac{\varphi }{2}\sin \frac{\Phi }{2}}{\Phi \sin \frac{\varphi
-\Phi }{2}}\right) .  \eqnum{38}
\end{equation}%
As follows from Eq. (35), the terms $d_{n-m}^{\alpha \beta }$ are
proportional to\ $\left( \frac{\omega _{2}}{\omega _{e}}\right) ^{2}$ and
can be neglected in Eq. (36). Therefore the energy levels are given by 
\begin{equation}
E_{m,n}^{\pm \left( \alpha \beta \right) }=\frac{1}{2}\left\{ \left(
m+n\right) \omega _{c}\pm \left[ \left( \omega _{e}-\Omega -\left(
m-n\right) \omega _{c}+\Delta _{3}\right) ^{2}+b_{n-m}^{2}\right]
^{1/2}\right\} .  \eqnum{39}
\end{equation}%
The normalized energy $\frac{2E_{m,n}^{\pm \left( \alpha \beta \right) }}{%
\omega _{e}}$ is presented in Fig.3 as a function of $\frac{\Omega }{\omega
_{e}}$ for $m-n=0.$

The average probability \cite{J.H.Shirley} of the resonance transition is 
\begin{equation}
\bar{P}_{\alpha \rightarrow \beta }=\frac{1}{2}\sum_{m,n}\frac{b_{m-n}^{2}}{%
\left( \omega _{e}-\Omega -\left( m-n\right) \omega _{c}+\Delta _{3}\right)
^{2}+b_{m-n}^{2}}.  \eqnum{40}
\end{equation}%
The probability calculated according to Eq. (40) is presented in Fig.4 at $%
\varphi =\frac{\pi }{2}$ as a function of $x=\frac{\Omega }{\omega _{e}}$
for the terms in the sum with $m-n=0,\pm 1,\pm 2$.

Using (40) the expression for the nuclear magnetization $M$, is the quantity
needed to compare with experiment, can be found: 
\begin{equation}
\frac{M}{M_{0}}=1-\sum_{m,n}\frac{b_{m-n}^{2}}{\left( \omega _{e}-\Omega
-\left( m-n\right) \omega _{c}+\Delta _{3}\right) ^{2}+b_{m-n}^{2}}. 
\eqnum{41}
\end{equation}%
Eqs. (40) and (41) coincide (by changing the summation index$:k=m-n$) with
the equations obtained for the transition probability and the magnetization
in the previous section (see Eqs. (23) and (24)).\ The magnetization
calculated according to Eq. (41) is presented in Fig.5.

{\bf Results and Discussion}

As follows clearly from the comparison of expressions (23), (24) , (40), and
(41), the calculations using both the semi-classical and Floquet methods
give the identical results for the transition probability and nuclear
magnetization. The results for the resonance conditions and for the shift $%
\Delta _{3}$of the resonant frequency obtained within the framework of these
methods are also identical. Moreover, the semi-classical approach allows one
to obtain automatically the shift $\Delta _{1}$ under off-resonance
conditions $\omega _{0}\neq \omega $ ($\Omega =0$) from RF excitation.
However the interpretation of the multi-frequency resonance transitions for
two-level energy system in the framework of the semi-classical approach\
involves difficulties.\ 

The Floquet method gives clear physical description of observed phenomena.
The time-independent Hamiltonian obtained using this method leads to the
multi-level energy spectrum. Each of these levels corresponds to a
stationary state of the system:\ spin dressed by photons. The coupling
between photons and spin allows one to interpret the resonances as
''anticrossing'' of the levels\ \cite{J.H.Shirley,Y.Zur}.

Matching of the computed results for the nuclear magnetization with the
experimental data displays series of essential differences. One of these
differences consists in the fact that the observed NMR\ signal does not damp
up to zero (Figs. 1 and 2), while the calculations predict zero
magnetization at the resonant frequency (Eq.(41) and Fig.5). A possible
reason of the observed finite value of $\frac{M}{M_{0}}$ is an inhomogeneity
of the RF field in the bulk of the sample resulting in a distribution of
angles $\varphi $ over the sample. To compare correctly the theoretical
result for the nuclear magnetization with the experiment, we have to take
into account the distribution of $\varphi $ over the real resonance line
with nonzero line width averaging expression (41) over the line: 
\begin{equation}
\left\langle \frac{M}{M_{0}}\right\rangle =\frac{1}{M_{0}}\int d\varkappa
M\left( \varkappa \right) g\left( \varkappa \right) .  \eqnum{42}
\end{equation}%
Let us consider the Lorentzian distribution of frequencies over a line
having width $c$ . It leads to the following distribution of angles $\varphi 
$%
\begin{equation}
g\left( \varkappa \right) =\frac{1}{\pi }\frac{c}{c^{2}+\varkappa ^{2}}, 
\eqnum{43}
\end{equation}%
where $\varkappa =\varphi -\varphi _{0}$ and $\varphi _{0}=\left\langle
\varphi \right\rangle _{V}$ is an angle averaged over the sample. After the
integration of (42) with the substitution of (43) we obtain 
\begin{equation}
\left\langle \frac{M}{M_{0}}\right\rangle =1-\sum_{n}\frac{1+a_{n}c}{%
a_{n}^{2}\left( 1-x-\frac{2\pi }{\varphi _{0}}n+\frac{\Delta _{3}t_{c}}{%
\varphi _{0}}\right) ^{2}+\left( 1+a_{n}c\right) ^{2}},  \eqnum{44}
\end{equation}%
where 
\begin{equation}
a_{n}=\frac{\varphi _{0}^{4}}{\left( \omega _{2}t_{c}\right) ^{2}\sin ^{2}%
\frac{\varphi _{0}}{2}}\left( 1+\frac{2\pi }{\varphi _{0}}n\right) ^{2}, 
\eqnum{45}
\end{equation}

The average magnetization, calculated according Eq. (44) with $c=0.018$, is
presented in Fig.6 at $\varphi _{0}=\frac{\pi }{2}$ as a function of $%
x=\left\vert \frac{\Omega }{\omega _{e}}\right\vert $ along with the
experimental results. Using the relationship $c=\left\langle \left( \varphi
-\varphi _{0}\right) /\varphi _{0}\right\rangle $ the dispersion of angles
can be estimated as $\left\langle \left( \varphi -\varphi _{0}\right)
\right\rangle =2.\,\allowbreak 827\,4\times 10^{-2}%
\mathop{\rm rad}%
$ or $1.\,\allowbreak 62^{0}$. One can see the excellent agreement of the
theoretical and experimental results.

The resonance transitions corresponding to negative frequencies $\Omega $
(see Fig. 5) can be interpreted as ones caused by the rotating in the
opposite direction component of the linearly polarized LF field .

The multi-frequency resonances were observed in the experiments with other
excitation schemes \cite{Y.Zur,Vega1986,Gromov} and explained by multiphoton
transitions between different states. \ Because the multiphoton transitions
involves intermediate states and the transition probability decreases with
number of photons \cite{Y.Zur,Vega1986,Gromov}, the intensity of the
resonance signal diminishes quickly with number of the mode. A peculiarity
of the excitation scheme proposed in the present paper is that, along with
multiphoton transitions, there are direct transitions between dressed spin
states with $\left( n-m\right) >1$ . Therefore, the transition probability
is the same for all the differences $n-m$ . As a result, the intensity of
the resonance signal is slowly decreased with number of the mode (Figs. 1
and 2).

{\bf Conclusions}

We have studied both experimentally and theoretically the dynamics of a spin
system with the spin 3/2 under simultaneous influence of DC, multiple-pulse
RF and LF fields. An important peculiarity of the applied excitation scheme
is the possibility of the measurement of the effective field $H_{e}$ and
observation of odd and even resonances with large mode numbers. It was shown
that the intensity of the resonance lines decreases slowly.

Both used theoretical approaches, the semi-classical and the Floquet
methods, give the identical results at the calculation of the resonant
frequencies, magnetization amplitudes and shifts of resonance frequency.
These results are in a good agreement with the experimental data for
different angles and mode numbers\ equaled to $n=0,\pm 1,\pm 2$ . \ Some
difference between the theoretical and experimental values of the signal
intensity can be explained by the inhomogeneity of the RF field over the
sample leading to the dispersion of the angle $\varphi $ .

The proposed technique can be incorporated for the study of the slow atomic
motion in systems involving spin-3/2 nuclei.

{\bf Acknowledgments}

This research was supported in part by a Grant from the U. S.-Israel
Binational Science Foundation (BSF).

\bigskip 

Figure captions:

Fig. 1 Time dependence of relative magnetization $\frac{M}{M_{0}}$ in the
absence of the LF field (open circle) and in the present of the LF field
(solid circle) at $\varphi =\frac{\pi }{2}$ in hexamethylbenzene $%
C_{6}(CH_{3})_{6}$.

Fig. 2 Dependence of relative magnetization $\frac{M}{M_{0}}$ on the LF
field frequency $\Omega $ at $\varphi =\frac{\pi }{2}$ in hexamethylbenzene $%
C_{6}(CH_{3})_{6}$.

Fig. 3 The normalized energy as a function of $x=\frac{\Omega }{\omega _{e}}$
for $m-n=0.$

Fig. 4 The probabilities as functions of $x=\frac{\Omega }{\omega _{e}}$ for 
$\varphi =\frac{\pi }{2}$ and $m-n=0,\pm 1,\pm 2$.

Fig. 5  {}Normalized magnetization $\frac{M}{M_{0}}$ as function of $%
x=\left\vert \Omega /\omega _{e}\right\vert $ calculated according Eq.(41)
for $\varphi =\frac{\pi }{2}$ and $m-n=0,\pm 1,\pm 2$.

Fig. 6 {}Dependence of averaged magnetization $\left\langle \frac{M}{M_{0}}%
\right\rangle $ of $x=\left\vert \Omega /\omega _{e}\right\vert $ at $%
\varphi _{0}=\frac{\pi }{2}$ in hexamethylbenzene $C_{6}(CH_{3})_{6}$: solid
curve - theory (44), open circle -experiment.   

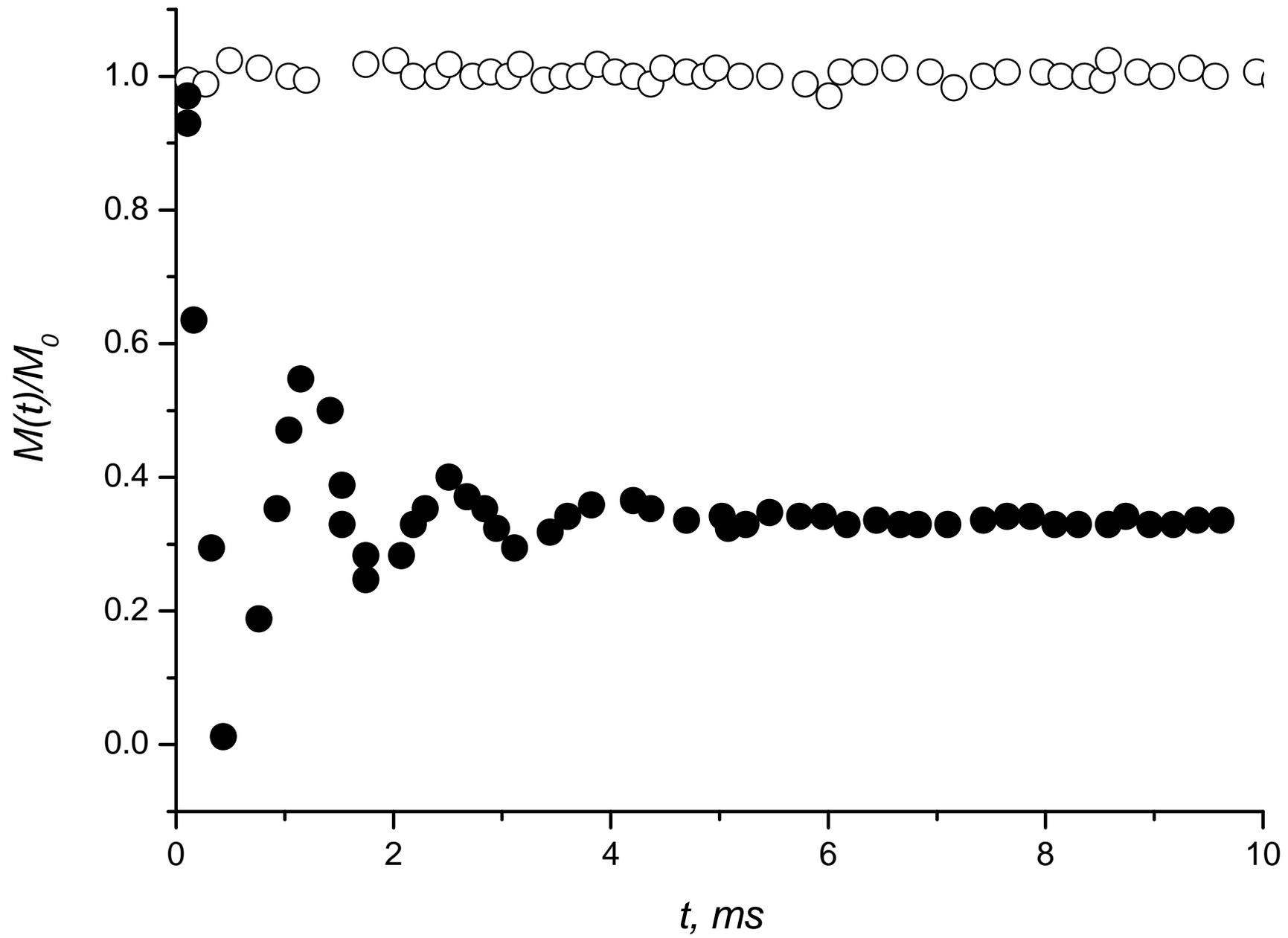
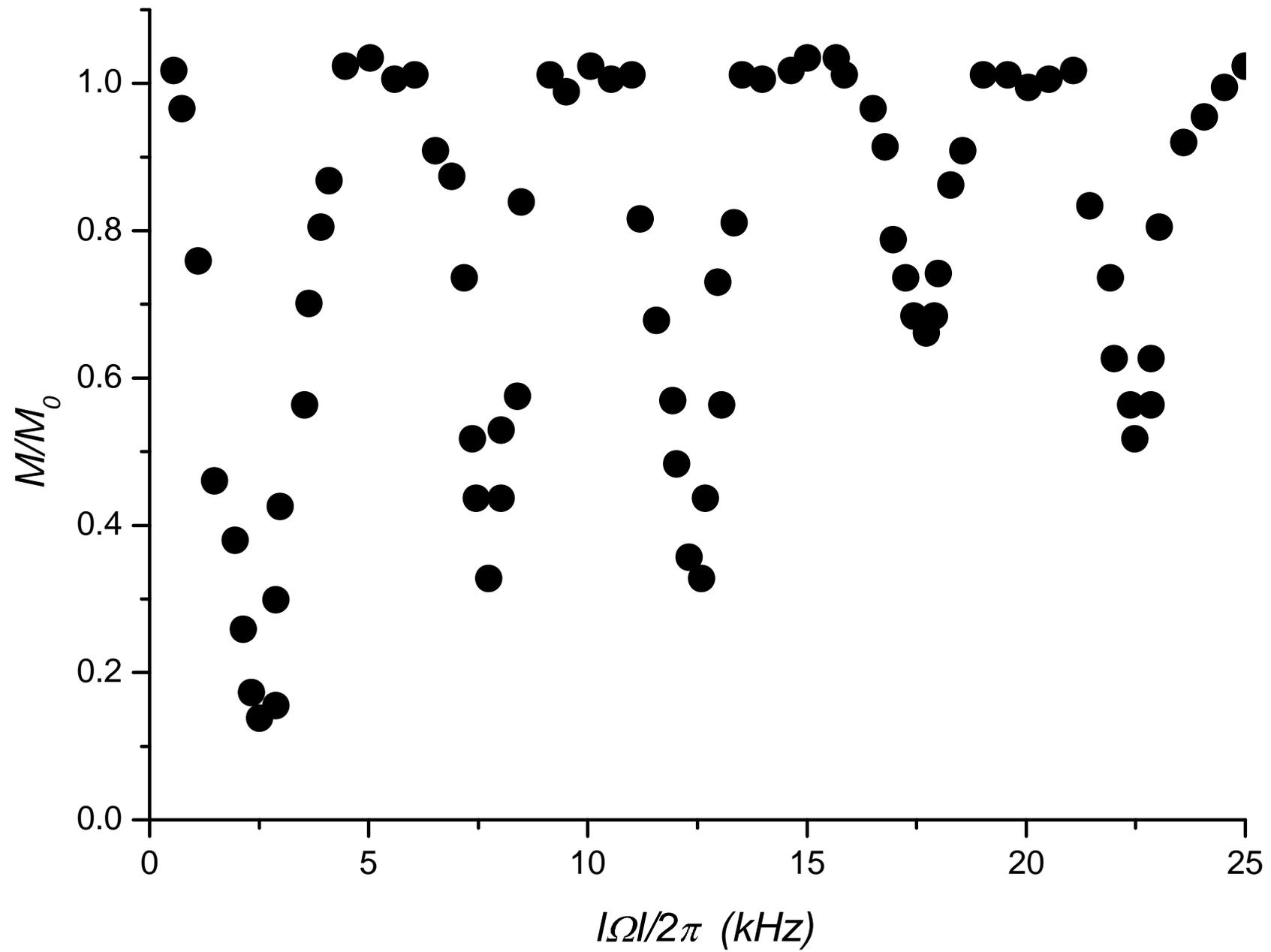
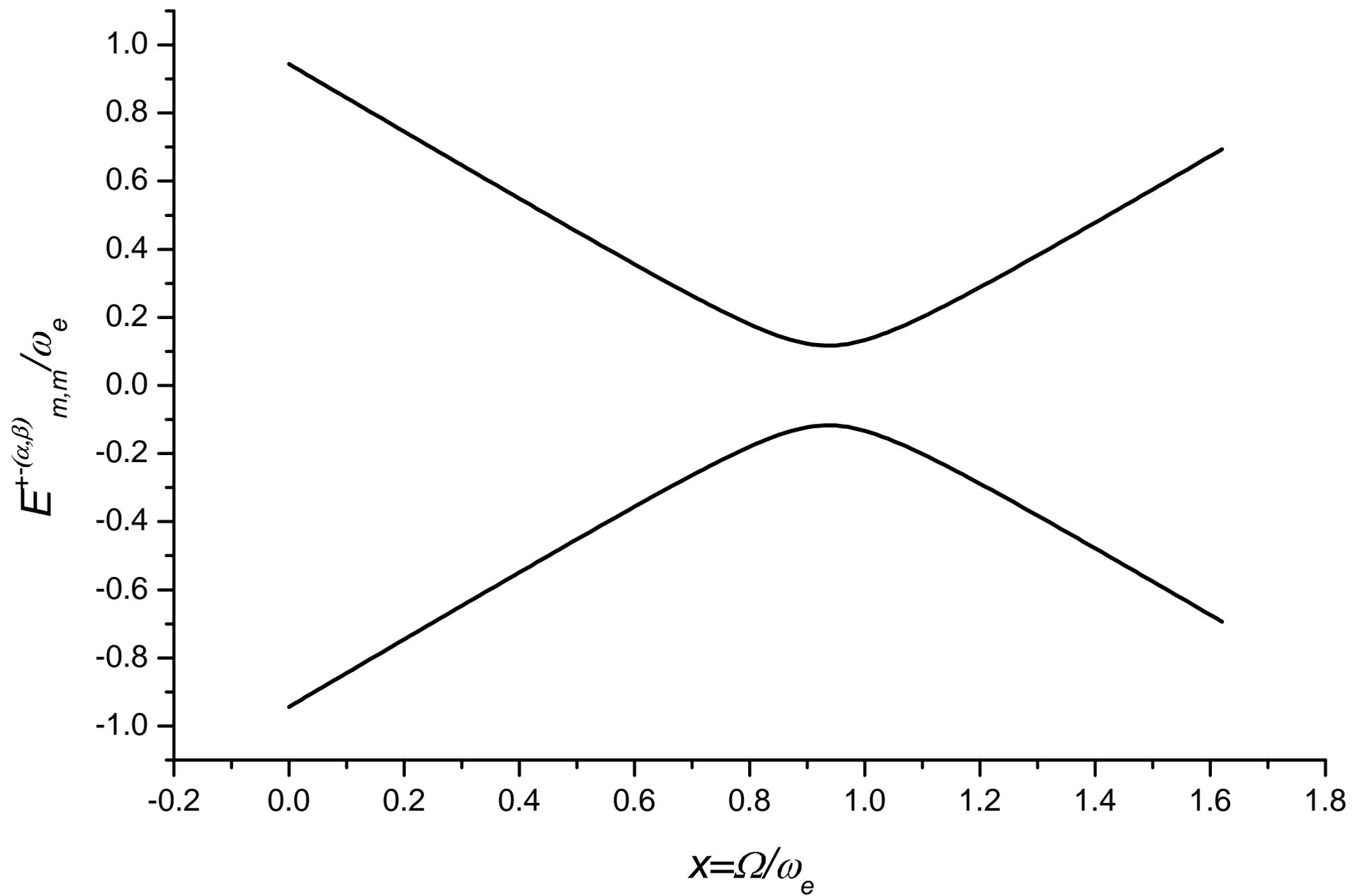
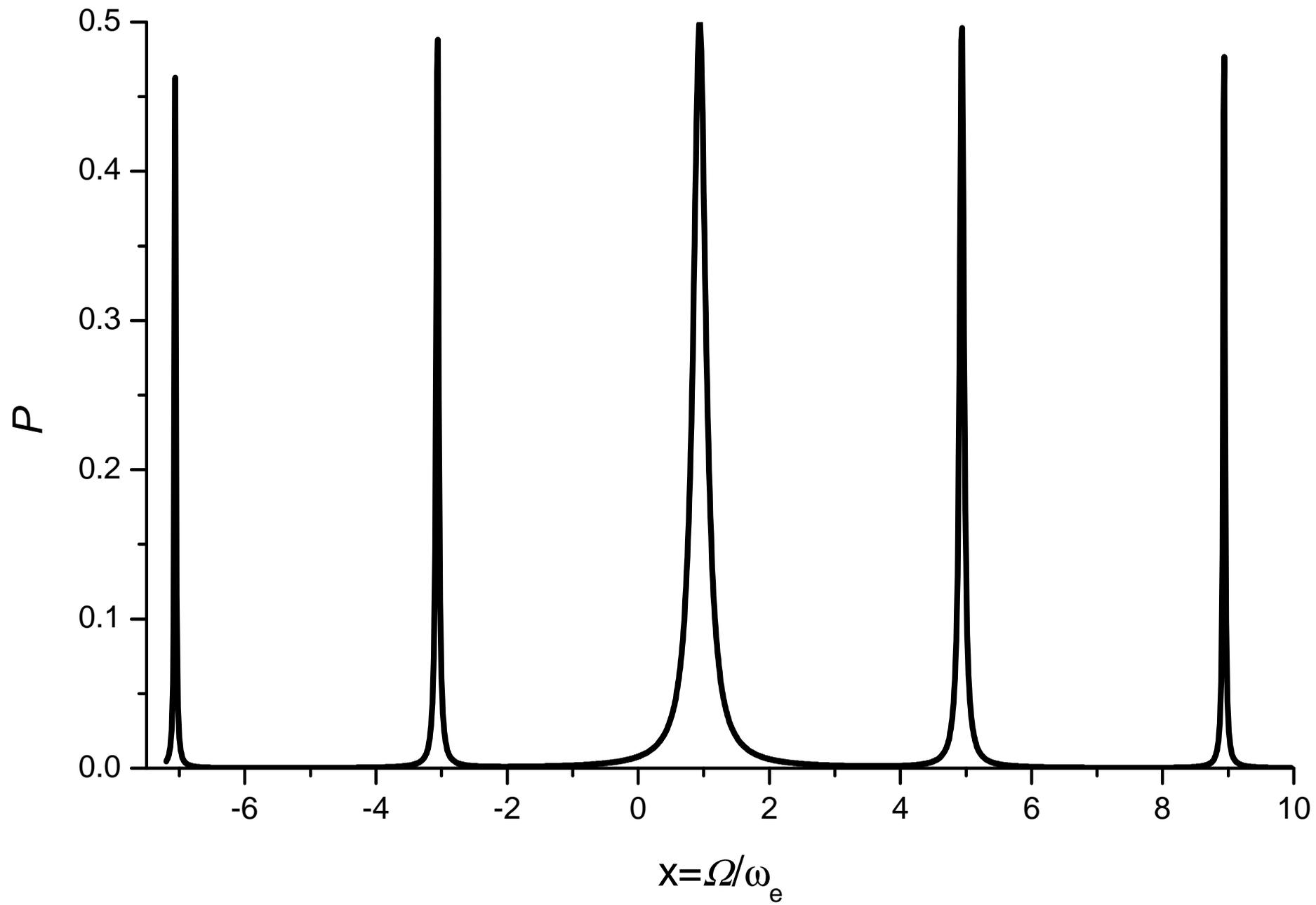
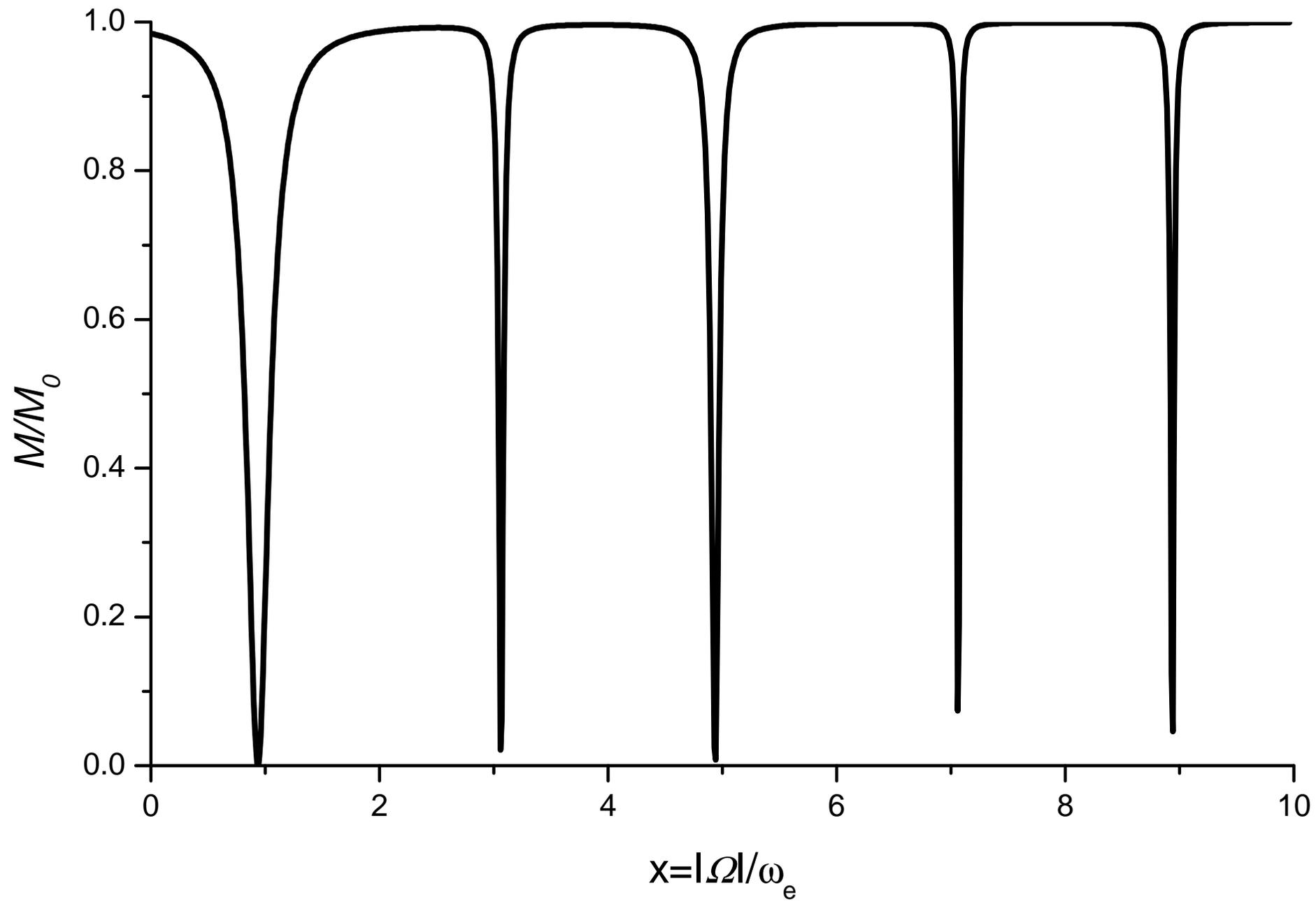
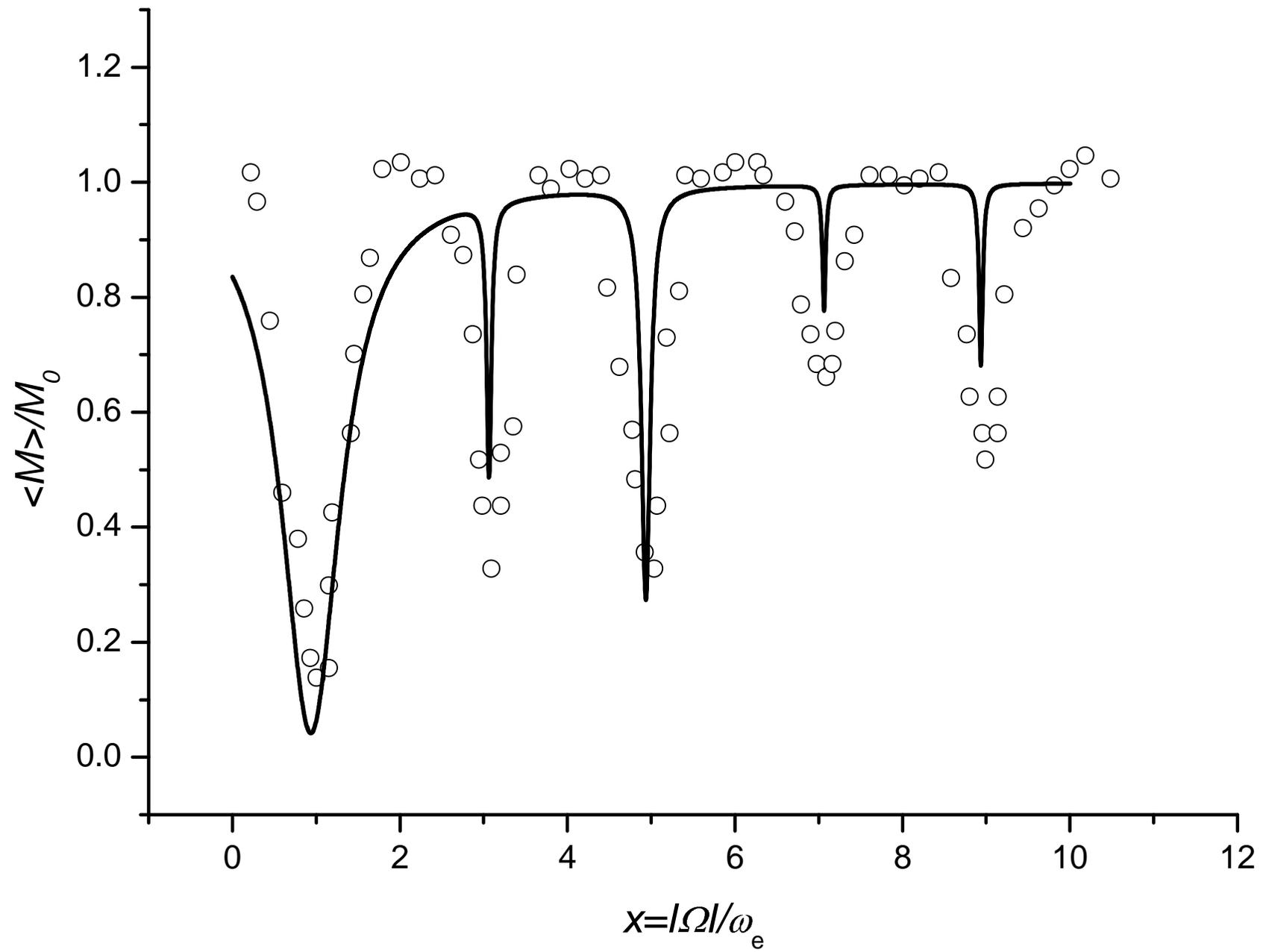
\end{document}